\documentclass{moriond}

\def\be{\begin{equation}}
\def\ee{\end{equation}}
\def\bea{\begin{eqnarray}}
\def\eea{\end{eqnarray}}

\newcommand{\Photo}{}

\begin{document}
\vspace*{4cm}
\title{Recent results from MAGIC
\vskip -0.1 cm}

\author{S. Mangano\\ 
on behalf of the MAGIC Collaboration
\vskip 0.0 cm}

\address{CIEMAT - Centro de Investigaciones Energ\'eticas, Medioambientales y Tecnol\'ogicas\\ 
Av. Complutense, 40, 28040 Madrid, Spain}

\maketitle\abstracts{
MAGIC is a system of two 17-m diameter Imaging Atmospheric Cherenkov 
Telescopes, located at an altitude of 2200 m in 
Roque de los Muchachos on the Canary island of La Palma,
exploring the gamma-ray sky above a few tens of GeV and 
up to tens of TeV. This system provides a low energy threshold 
and a fast automated response to transient phenomena.
In this contribution, some selected results of MAGIC, 
which has been collecting data for more than 20 years, are reviewed.
Special attention is given to multiwavelength and multimessenger astronomy, 
such as GRB 201216C, 
the farthest ground-based detection of a very-high-energy gamma-ray bursts, 
as well as the RS Ophiuchi nova.
The scientific program also includes measuring the 
cosmic-ray electron positron spectrum, 
estimating the size of stars using intensity interferometry, 
studying gravitational lensing and searching for dark matter 
in spheroidal galaxies.
Finally, a glimpse into the future is given by 
presenting the performance of the joint observations 
with the first Large-Sized Telescope from the 
Cherenkov Telescope Array and MAGIC.}

\section{Introduction}
MAGIC~\cite{magic} (Major Atmospheric Gamma-ray Imaging Cherenkov) is 
situated 2200 meters above sea level at the Roque de los Muchachos 
Observatory on the Canary Island of La Palma. 
In 2004, data collection began with the first telescope, which had a diameter of 17 meters and a second telescope was added in 2009 in order to significantly increase the sensitivity through stereo observations~\cite{magicupgrad}. 
Later, further hardware upgrades and software improvements helped to gain scientific performance~\cite{magiczenith,magicsum}.
With H.E.S.S.~\cite{hess} and VERITAS~\cite{veritas}, MAGIC is among the 
most sensitive instrument for high-energy gamma-ray astrophysics between a few tens of GeV and up to tens of TeV.
The MAGIC performance parameters~\cite{magicperform} are a field of view of up to $3.5^{\circ}$, a duty cycle of around 15\%, an angular resolution at energies of a few hundred GeV of less than $0.07^{\circ}$, and an energy resolution of around 16\%. The main objective of the MAGIC Collaboration is to explore the most energetic phenomena in astrophysics by detecting high-energy photons generated by cosmic accelerators.
MAGIC has discovered sources belonging to several different types of such cosmic accelerators, such as binary systems, pulsars, pulsar wind nebulae, supernova remnants, active galactic nuclei, and gamma-ray bursts, helping to improve our knowledge of these energetic phenomena.

\section{MAGIC discoveries}
To increase the MAGIC discovery potential and to make most 
of its limited observation time and field of view, alerts 
from electromagnetic, neutrino and gravitational-wave observatories trigger immediate follow-up observations in both multiwavelength 
(see subsections 2.1, 2.3 and 2.4) and multimessenger (see subsection 2.2) contexts.

\subsection{Gravitationally-lensed blazar QSO B0218+357
\vskip -0.1 cm}
The theory of general relativity predicts that compact objects would cause space to be curved. Gravitational lensing is the phenomena that results from propagation of light rays in this curved space. When a compact object is situated along the line of sight between the observer and the source, a gravitational lens magnifies the source and creates several duplicated images of the original object. Due to different geometrical paths and relativistic time delays the emission arrive at different times in the various images. The blazar~\footnote{A blazar is a type of active galactic nucleus with a jet of particles and energy that points towards the Earth.} QSO B0218+357 at redshift z=0.94 was the first gravitationally-lensed blazar observed in the very-high-energy (VHE, \mbox{E$~> 100$ GeV}) gamma-ray range.  The angular resolution of MAGIC does not allow to spatially resolve the images, but the expected time delay difference allowed the detection of delayed component of emission. Due to the fact that these photons travel cosmological distances through diffuse extragalactic radiation fields, they can be used to compute the gamma-ray attenuation caused by the existing extragalactic background light~\cite{glb0218}. A multiwavelength campaign from 2016 to 2020 has provided a better understanding of the emission from QSO B0218+357, but unfortunately, no VHE gamma-ray flares have been observed. However the radio data have been used to improve the lens modeling to evaluate image magnifications and time delays~\cite{glb0218_2022}. Moreover, the multiwavelength broadband spectral energy distribution is fitted with a two zone synchroton self Compton and external Compton model as shown in the left panel of Figure 1. 

\begin{figure}
\begin{minipage}{0.5\linewidth}
\centerline{\includegraphics[width=1.1\linewidth]{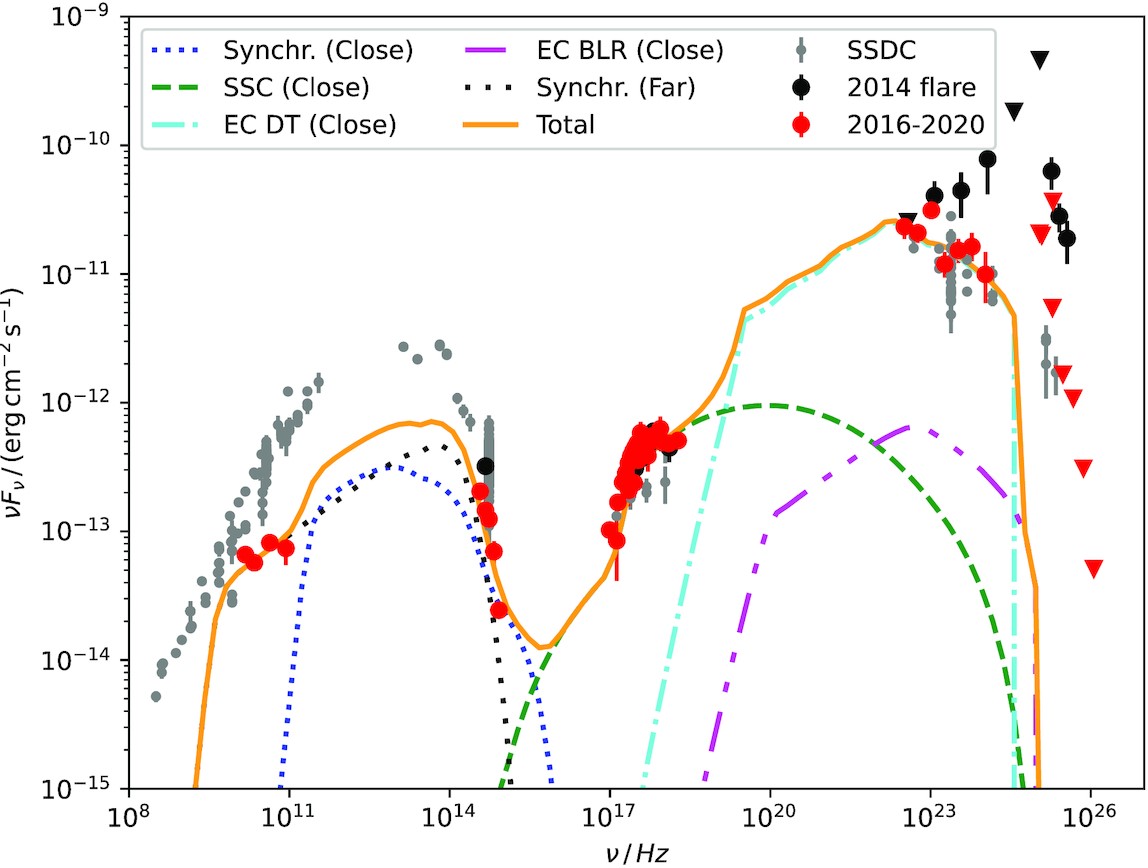}}
\end{minipage}
\hfill
\begin{minipage}{0.5\linewidth}
\centerline{\includegraphics[width=0.85\linewidth]{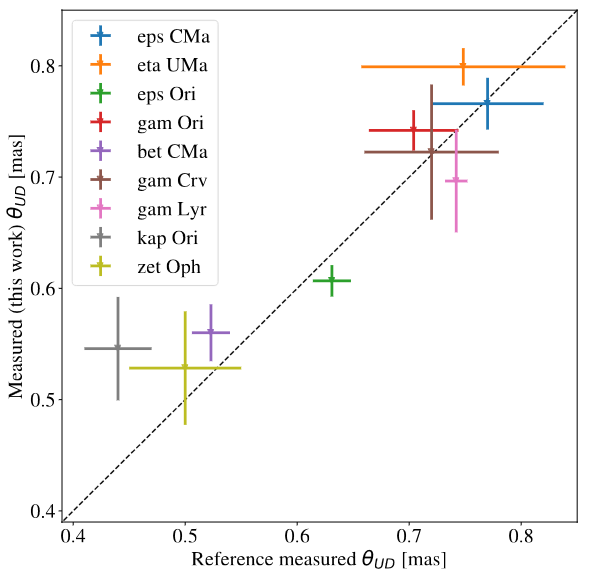}}
\end{minipage}
\caption[]{Left panel: Multiwavelength broadband spectral energy distribution of the gravitationally-lensed blazar QSO B0218+357 composed of contemporaneous data to the MAGIC observations in 2016~-~2020 (red points), historical data (gray) and data from the 2014 flare~\cite{glb0218} (in black) compared with a two zone synchrotron self Compton (SSC) and external Component (EC) model. The total emission is shown with an orange line, which is composed of synchrotron emission (dotted curves) and SSC emission (dashed curves). The model takes into account the lensing magnification, the absorption in the lens galaxy, and the extragalactic background light attenuation. The left panel is reprinted from Reference ~\cite{glb0218_2022} (p. 2358). Right panel: MAGIC measured angular stellar diameter (in microarcseconds) of nine stars compared with existing reference diameter measurements over similar wavelengths. The right panel is reprinted from Reference ~\cite{secondII} \mbox{(p. 4397)}.}
\label{fig:posiblelayout}
\end{figure}

\subsection{Multimessenger astrophysics with blazar TXS 0506+056
\vskip -0.1 cm}
Multimessenger astrophysics aims for simultaneous observations of 
gamma-rays, neutrinos, gravi-tational-waves, and cosmic-rays. 
Such combinations are expected to 
give information on the acceleration mechanism of astrophysical sources.  
The blazar TXS 0506+056 received much attention in 2017 when MAGIC observed an enhanced flux of gamma-rays at the same time as IceCube detected a high-energy neutrino from the same object. This intriguing VHE gamma-ray flare and high-energy neutrino association is still the most significant observed one~\cite{txs1}. Before this event, TXS 0506+056 was relatively poorly studied. In order to better understand this source, a multiwavelength campaign spanning 16 months (from November 2017 to February 2019) was organized. During this time period the source showed its lowest VHE gamma-ray emission and was in an enhanced flux state in December 2018. However during this flaring activity there was no additional neutrino detected from this source direction~\cite{txs2}. These results show there exist complex processes in the emission of blazars and evidences the importance of multiwavelength and multimessenger observations in better understanding their behavior.

\subsection{Gamma-ray bursts
\vskip -0.1 cm}
Gamma-ray bursts (GRBs) are highly explosive events 
that occur in distant galaxies, emitting large amounts of 
energy as electromagnetic radiation across the full energy range.
Thanks to the fast repointing and low energy capability, MAGIC has detected several GRBs in the last few years at TeV energies, after a long period for search of such emission from GRBs. The GRB 190114C is the first GRB from which VHE emission was claimed, whereas the GRB 201216C at z=1.1 is the farthest ground-based detected VHE GRB. Both GRBs were accompanied by a rich multiwavelength campaign, and further details can be found in References~\cite{GRBfirst,GRBsecond,GRB}.  

\subsection{Outburst of recurrent nova RS Ophiuchi
\vskip -0.1 cm}
MAGIC observed the first VHE gamma-ray recurrent nova, RS Ophiuchi, revealing that protons are being accelerated to high energies in the nova shock~\cite{rsoph}. This suggests that hadronic \mbox{acceleration} plays an important role in the energy release mechanisms of these objects. Furthermore, these accelerated protons will sooner or later escape the nova shock and (slightly) contribute to the cosmic-rays that constantly bombard Earth. 

\subsection{Cosmic-ray electrons and positrons
\vskip -0.1 cm}
The cosmic-ray electrons and positrons (CREs) make up less than 1\% of the total cosmic-ray flux, composed mainly of cosmic-ray protons. Due to the low mass of the electrons and positrons, they experience significant radiative losses and these losses increase quickly with increasing particle energy. 
As a result, the tortuous trajectory in the Galactic magnetic field of these TeV CREs are expected to reach a distance from the source of less than 1 kiloparsec.
Therefore such a CREs measurement gives a good indication of existing nearby sources. MAGIC has used machine learning classification to distinguish CREs from protons and heavier nuclei. By analyzing 220 hours of MAGIC data, a CRE spectrum spanning an energy range from 300 GeV to 6 TeV has been measured~\cite{Yating} and aligns well with the results from other experiments, like H.E.S.S.~\cite{hess2009}, VERITAS~\cite{veritas2018}, DAMPE~\cite{at,Dampe}, AMS~\cite{AMS2021} and CALET~\cite{Calet2019}.

\subsection{Optical intensity interferometry
\vskip -0.1 cm}
The optical intensity interferometry observations give the possibility to measure the apparent angular diameter of a star using the so-called Hanbury Brown and Twiss effect~\cite{bt1,bt2}.  After a first successful proof of concept campaign with minimal hardware changes on the MAGIC telescopes in 2019~\cite{firstII}, further hardware and software improvements have been introduced over the last few years. These updates made it possible to measure the diameter of 22 stars~\cite{secondII}, of which nine correspond to stars whose diameter was known from other measurements. The right panel of \mbox{Figure 1} shows good agreement between the MAGIC measured diameters with previous observations from other experiments at similar wavelengths.

\section{Conclusion}
Unfortunately, in this proceeding there was not enough room 
to discuss a number of new
and exciting results which were illustrated 
at the $58^{th}$ Rencontres de Moriond, 
such as the the MAGIC detection and interpretation of the 
blazar B2 1811+31~\cite{dc,lo}, 
the search for dark matter annihilation with a combined analysis of dwarf spheroidal galaxies from five gamma-ray experiments~\cite{dk,ca}, the search for line-like features from dark matter in the Galactic Center region~\cite{ti,gammalines} as well as the performance of the joint observations with MAGIC and the first Large-Sized Telescope~\cite{LST} from the Cherenkov Telescope Array~\cite{CTAMangano,CTAweb}. The MAGIC telescopes are still being developed and maintained after 20 years of data collection, therefore it is anticipated that they will
provide further important scientific data in the coming years as the Cherenkov Telescope Array is constructed.

\section*{Acknowledgments
\vskip -0.25 cm}
The author expresses gratitude to the Instituto de Astrof\'isica de Canarias for the excellent working conditions at the Roque de los Muchachos Observatory in La Palma and to the Spanish Ministry of Science and Innovation and the Spanish Research State Agency through the grant PID2022-138172NB-C41 for the financial support.



\section*{References


\vskip -0.25 cm}
\vspace*{0mm}
\begin{thebibliography}{99}
\bibitem{magic} MAGIC website: ~https://magic.mpp.mpg.de/.
\bibitem{magicupgrad} J. Aleksi\'c et al., Astropart. Phys., 72 (2016) 61-75.
\bibitem{magiczenith} V. A. Acciari et al., A \& A 635 (2020) A158.
\bibitem{magicsum} F. Dazzi et al., IEEE Transactions on Nuclear Science, 68 (2021) 7. 
\bibitem{hess} H.E.S.S. website: ~https://www.mpi-hd.mpg.de/hfm/HESS/.
\bibitem{veritas} VERITAS website: ~https://veritas.sao.arizona.edu/.

\bibitem{magicperform} J. Aleksi\'c et al.,  Astropart. Phys., 72 (2016) 76-94.

\bibitem{glb0218} M. L. Ahnen et al., A \& A 595 (2016) A98.
\bibitem{glb0218_2022} V. A. Acciari et al., MNRAS 510 (2022) 2, 2344-2362.
\bibitem{txs1} S. Ansoldi et al., ApJL 863 (2018) L10.
\bibitem{txs2} V. A. Acciari et al., ApJ 927 (2022) 197.
\bibitem{GRBfirst} V. A. Acciari et al., Nature 575 (2019) 455-458.
\bibitem{GRBsecond} V. A. Acciari et al., Nature 575 (2019) 459-463. 
\bibitem{GRB} H. Abe et al., MNRAS 527 (2024) 3, 5856-5867.
\bibitem{rsoph} V. A. Acciari et al., Nature Astronomy 6 (2022) 689-697.

\bibitem{Yating} Y. Chai et al., PoS (ICRC2023) 323.
\bibitem{hess2009} F. Aharonian et al., A \& A 508 (2009) 561-564.
\bibitem{veritas2018} A. Archer et al., PRD 98 (2018) 062004.
\bibitem{at} A. Tykhonov, Proceedings of Rencontres de Moriond, VHEPU 2024.
\bibitem{Dampe} G. Ambrosi et al., Nature 552 (2017) 63-66.
\bibitem{AMS2021} M. Aguilar et al., Physics Reports 894 (2021) 1-116.
\bibitem{Calet2019} O. Adriani et al., PRL 120 (2018) 261102.

\bibitem{bt1} R. Hanbury-Brown, R. Q. Twiss, Phil. Mag. 45 (1954) 663-682.
\bibitem{bt2} R. Hanbury-Brown, R. Q. Twiss, Nature 178 (1956) 1046-1048.


\bibitem{firstII} V. A. Acciari et al., MNRAS 491 (2020) 2, 1540-1547.
\bibitem{secondII} S. Abe et al., MNRAS 529 (2024) 4, 4387-4404.

\bibitem{dc} D. Cerasole, Proceedings of Rencontres de Moriond, VHEPU 2024.
\bibitem{lo} S. Loporchio et al., PoS (ICRC 2023) 725.
\bibitem{dk} D. Kerszberg, Proceedings of Rencontres de Moriond, VHEPU 2024.
\bibitem{ca} C. Armand et al., PoS (ICRC 2021) 528.
\bibitem{ti} T. Inada, Proceedings of Rencontres de Moriond, VHEPU 2024.
\bibitem{gammalines} H. Abe et al., PRL 130 (2023) 061002.
\bibitem{LST} H. Abe et al., A \& A, 680 (2023) A66.
\bibitem{CTAMangano} S. Mangano, arXiv:1705.07805, Proceedings of Rencontres de Moriond, VHEPU 2017.
\bibitem{CTAweb} CTA website: ~https://www.cta-observatory.org/.





\end{thebibliography}
\end{document}